\documentclass[%
aps,
prl,
reprint,
superscriptaddress,
nofootinbib,
nobibnotes,
floatfix,
]{revtex4-2}

\usepackage{xcolor}
\usepackage{graphicx}
\graphicspath{{figln/}{fig/}}
\usepackage{amsmath,amssymb,amsfonts,mathrsfs,bm}
\usepackage[colorlinks=true,linkcolor=blue,citecolor=blue]{hyperref}

\usepackage{iftex}
\ifpdftex
    \usepackage{CJKutf8}
    \newcommand{\Ckai}[1]{\begin{CJK*}{UTF8}{gkai}#1\end{CJK*}}
\fi
\ifxetex
    \usepackage{xeCJK}
    \newcommand{\Ckai}[1]{\textit{#1}}
\fi

\newcommand{\dd}{\mathrm{d}}
\newcommand{\ee}{\mathrm{e}}
\newcommand{\ii}{\mathrm{i}}

\newcommand{\Lie}{\mathcal{L}}
\newcommand{\Rot}{\mathcal{R}}
\newcommand{\avg}[1]{\langle #1 \rangle}
\newcommand{\Imo}{\operatorname{Im}}

\begin{document}

\title{%
Cyclic Boson Stars with Quantum Vortex Molecules
}

\author{Xing-Yu Yang (\Ckai{杨星宇})}
\email[Corresponding author:~]{xingyuyang@kias.re.kr}
\affiliation{Quantum Universe Center (QUC), Korea Institute for Advanced Study, Seoul 02455, Republic of Korea}

\author{Rong-Gen Cai (\Ckai{蔡荣根})}
\email[Corresponding author:~]{caironggen@nbu.edu.cn}
\affiliation{Institute of Fundamental Physics and Quantum Technology, Ningbo University, Ningbo 315211, China}
\affiliation{Zhejiang Key Lab of Extreme Universe, Ningbo University, Ningbo 315211, China}

\begin{abstract}
    We find a new type of stationary boson star with cyclic rotational symmetry in the minimal model of complex scalar field coupled to gravity. Unlike conventional axisymmetric spinning boson stars, these configurations can carry noninteger angular momentum per particle in units of the reduced Planck constant, revealing a new way for boson stars to carry angular momentum. We identify the vortex assemblies within these stars as self-gravitating quantum vortex molecules, in which gravity supports a stationary vortex arrangement despite internal circulation. These structures provide a relativistic realization of quantum vortex molecules studied in condensates and motivate possible laboratory analogues.
\end{abstract}

\maketitle

\textit{Introduction---}%
Boson stars are localized configurations of bosonic fields bound by their own gravity and are often interpreted as macroscopic Bose--Einstein condensates~\cite{Kaup:1968zz,Ruffini:1969qy}.
They connect coherent quantum matter with the nonlinear dynamics of gravity and motivate applications ranging from particle physics to astrophysics and cosmology~\cite{Jetzer:1991jr,Schunck:2003kk,Liebling:2012fv,Visinelli:2021uve,CalderonBustillo:2020fyi,Evstafyeva:2024qvp,Lira:2024cma,Banik:2025xwl,Yang:2025yej}.
Understanding their possible forms is therefore essential for exploring the gravitational manifestations of bosonic matter.

Conventional stationary spinning boson stars are constructed under the assumption of axisymmetry and have integer angular momentum per particle in units of the reduced Planck constant~\cite{Yoshida:1997qf,Schunck:1996he,Grandclement:2014msa}, whereas static multipolar boson stars with discrete spatial symmetries have vanishing total angular momentum~\cite{Herdeiro:2020kvf}.
Here we find a new type of stationary boson star with cyclic rotational symmetry in the extensively studied minimal model of a single complex scalar field coupled to gravity.
Their geometry and matter distribution remain stationary with only discrete rotational symmetry about the spin axis, while their angular momentum per particle can take noninteger values, revealing a new way for boson stars to carry angular momentum.

Quantum vortex molecules are organized assemblies of quantized vortices studied in Bose--Einstein condensates~\cite{Kasamatsu:2004tvg,2008PhRvA..78e3610G,Eto:2012rc,Eto:2013spa}.
We find that the vortices within cyclic boson stars form such molecules, with gravity binding the condensate and supporting a stationary cyclic arrangement despite internal circulation.
These structures provide a relativistic realization of quantum vortex molecules without external confinement and motivate possible laboratory analogues.

\textit{Cyclic boson stars---}%
We set $c=\hbar=G=1$ and use the metric signature $(-+++)$.
A free complex scalar field $\phi$ of mass $\mu$ minimally coupled to Einstein gravity is described by the action
\begin{equation}
    I=\int\dd^4x\sqrt{-g}\left[
        \frac{R}{16\pi}
        -\frac12\left(\nabla_a\bar{\phi}\nabla^a\phi+\mu^2|\phi|^2\right)
    \right],
    \label{eq:action}%
\end{equation}
which yields the Einstein--Klein--Gordon (EKG) equations
\begin{subequations}
    \begin{align}
        G_{ab}&=8\pi T_{ab},
        \\
        \nabla_a\nabla^a\phi&=\mu^2\phi.
    \end{align}
    \label{eq:EKG}%
\end{subequations}
Here $G_{ab}$ is the Einstein tensor, $T_{ab}$ is the scalar stress-energy tensor, and $\nabla_a$ is the covariant derivative associated with the metric $g_{ab}$.
We seek regular, horizonless, asymptotically flat solutions with a localized scalar field and stationary geometry.

Let $\bm{J}\ne0$ be the Arnowitt--Deser--Misner (ADM) angular momentum in the asymptotic rest frame~\cite{Regge:1974zd,Gourgoulhon:2012ffd}.
Nonzero angular momentum selects a direction but does not require axisymmetry.
Any rotational isometry preserving the stationary structure and time orientation induces a proper rotation $\Rot$ at infinity satisfying $\bm{J}=\Rot\bm{J}$.
Such rotations fix the spin direction and form a subgroup of $\mathrm{SO}(2)$, whose closed subgroups are $\mathrm{SO}(2)$ itself and the cyclic groups $C_q$.
Cyclic symmetry is therefore the natural discrete alternative to axisymmetry for a spinning configuration.
We seek solutions with integer $q\ge2$, since $C_1$ denotes the trivial group.

Choosing the cyclic axis as the $z$ axis, we impose
\begin{subequations}
    \begin{align}
        &\Lie_tg_{ab}=0,
        \quad
        \phi=\ee^{-\ii\omega t}\psi(r,\theta,\varphi),
        \label{eq:stationary}%
        \\
        &\Rot_q^*g_{ab}=g_{ab},
        \quad
        \psi(r,\theta,\varphi+2\pi/q)=\ee^{2\pi\ii k/q}\psi(r,\theta,\varphi).
        \label{eq:cyclic}%
    \end{align}
\end{subequations}
Here $\Lie_t$ denotes the Lie derivative along the stationary Killing field, $\Rot_q^*$ denotes the pullback by the rotation $\Rot_q:\varphi\to\varphi+2\pi/q$, and the integer $k$ is defined modulo $q$, with $0\le k<q$.
The common temporal phase of the scalar field makes the stress-energy tensor time-independent.
The scalar can acquire a constant phase under a cyclic rotation because this phase cancels from the stress-energy tensor.
The cyclic condition restricts the azimuthal expansion to
\begin{equation}
    \psi(r,\theta,\varphi)
    =\sum_{s\in\mathbb Z}\psi_s(r,\theta)\ee^{\ii(k+sq)\varphi}.
    \label{eq:psi}%
\end{equation}
Conventional axisymmetric stars are included as single-mode solutions with azimuthal number $m_s=k+sq$.
Multiple occupied modes permit cyclic configurations with stationary interference patterns.
Their amplitudes and relative phases must be determined self-consistently with the metric by the nonlinear EKG equations.

We develop a systematic method for constructing cyclic solutions of the EKG equations, using solutions of their weak-field limit to initialize the relativistic calculation.
Define $\kappa=\sqrt{1-\omega^2/\mu^2}$ with $0<\omega<\mu$ and introduce dimensionless rescaled coordinates $(T,\bm{X})=\mu\kappa (t,\bm{x})$ with $R=|\bm{X}|$.
In the limit $\kappa\to0$, the EKG equations reduce to the Schr\"odinger--Poisson (SP) equations~\cite{Liebling:2012fv}
\begin{subequations}
    \begin{align}
        (-\Delta_{\bm{X}}+U)\Psi&=-\Psi,
        \\
        \Delta_{\bm{X}} U&=|\Psi|^2,
    \end{align}
    \label{eq:SP}%
\end{subequations}
where $\Psi(\bm{X})=\sqrt{8\pi}\,\psi(\bm{x})/\kappa^2$ is the rescaled scalar field and $U$ is the rescaled gravitational potential, related to the Newtonian potential by $U_{\mathrm{N}}=\kappa^2U/2$.

We obtain the SP configurations by continuation from a spherically averaged problem using the auxiliary system
\begin{subequations}
    \begin{align}
        [-\Delta_{\bm{X}}+\avg{U_h}+h(U_h-\avg{U_h})]\Psi_h&=-\Psi_h,
        \\
        \Delta_{\bm{X}} U_h&=|\Psi_h|^2.
    \end{align}
    \label{eq:SP_h}%
\end{subequations}
Here $\avg{U_h}(R)=(4\pi)^{-1}\int U_h(R,\Omega)\dd\Omega$, and $0\le h\le1$ interpolates between the averaged problem at $h=0$ and the physical SP equations~\eqref{eq:SP} at $h=1$.
At $h=0$, we construct seeds from solutions of the spherically averaged problem of the form $\Psi_0(R,\theta,\varphi)=R_{n\ell}(R)\sum_m c_mY_{\ell m}(\theta,\varphi)$, with $|m|\le\ell$, $m\equiv k\pmod q$, and $\sum_m|c_m|^2=1$.
Here $n$ counts the radial nodes of $R_{n\ell}$, and $\ell$ is the seed's spherical-harmonic degree.
For fixed $(n,\ell)$, all normalized mixtures share the same radial problem, whose solution with regularity at the origin and decay at infinity gives $R_{n\ell}(R)$.
A single azimuthal mode is axisymmetric, so cyclic seeds must mix at least two allowed modes; the lowest-degree pair, $m=k$ and $m=k-q$, gives $\ell\ge\max(k,q-k)$.
We determine $c_m$ by requiring the nonspherical gravitational interaction energy $\widetilde W(c)=\frac12\int|\Psi_0|^2(U_0-\avg{U_0})\dd^3X$ to be stationary under variations of $c_m$, $\delta\widetilde W(c)=0$.
Here $U_0$ solves $\Delta_{\bm X}U_0=|\Psi_0|^2$, with $U_0\to0$ at infinity.
This condition is necessary for smooth continuation and follows from the field equations at first order in $h$.
The radial profile $R_{n\ell}(R)$ and coefficients $c_m$ together determine a seed $\Psi_0$.
This construction can therefore yield distinct seeds at the same $(n,\ell)$, corresponding to different constrained stationary points of $\widetilde W(c)$.

We use the resulting seed $\Psi_0$ to initialize numerical continuation in $h$.
Branches reaching $h=1$ yield SP solutions, which we use as initial approximations for solving the EKG equations at small but finite $\kappa$.
We then continue these EKG solutions in $\kappa$ to obtain cyclic boson-star families, retaining all metric components and both the real and imaginary parts of the scalar field.
The final fields allow all harmonics compatible with cyclic covariance within the numerical cutoffs; the seed labels do not restrict their angular degree or radial profile.

\begin{figure*}[htbp]
    \centering
    \includegraphics[]{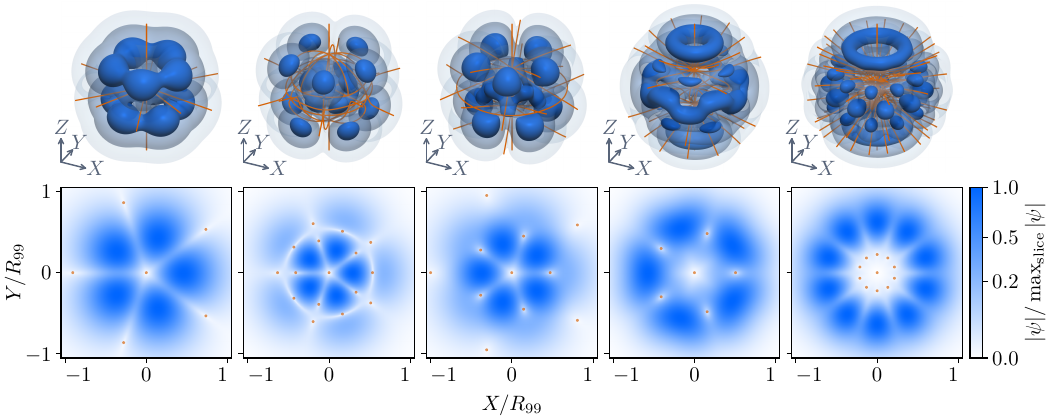}
    \caption{%
    Five cyclic boson stars in the sector $(k,q)=(2,5)$ at a common reduced charge $\mu^2N=0.1$.
    From left to right, the seed labels $(n,\ell)$ are $(0,3)$, $(1,3)$, $(0,4)$, $(0,7)$, and $(0,7)$; the fractional bindings are $1-M/(\mu N)\simeq(5.69,\,2.80,\,3.85,\,1.59,\,1.57)\times10^{-5}$; and the angular momenta per particle are $J/N\simeq(1.601,\,0.2514,\,-1.044,\,3.927,\,-0.00905)$.
    Top: scalar-amplitude isosurfaces at $|\psi|/\max|\psi|=0.05$, $0.2$, and $0.55$.
    Bottom: slices at $Z=0.1R_{99}$, each normalized by its maximum in the displayed slice.
    Lengths are scaled by each solution's $R_{99}$, the coordinate radius enclosing $99\%$ of the mass.
    Orange lines show resolved vortex filaments clipped to $R\le R_{99}$; orange points mark where vortices cross the displayed slice.
    }
    \label{fig:amplitude}
\end{figure*}

The global $\mathrm{U}(1)$ symmetry gives a conserved Noether charge $N$, interpreted as the particle number~\cite{Liebling:2012fv,Visinelli:2021uve}.
As concrete examples, Fig.~\ref{fig:amplitude} shows five cyclic boson-star solutions of the EKG equations in sector $(k,q)=(2,5)$ at a common reduced charge $\mu^2N=0.1$, corresponding to distinct values of $\kappa$ in the range $0.0097\lesssim\kappa\lesssim0.0185$.
These are distinct configurations of self-gravitating Bose--Einstein condensates with the same particle number.
The first three configurations originate from seeds with respective $(n,\ell)$ labels $(0,3)$, $(1,3)$, and $(0,4)$.
The last two configurations originate from distinct seeds with the same labels $(n,\ell)=(0,7)$.
All five configurations have cyclic symmetry, with distinct lobe arrangements and radial structures in their scalar-amplitude distributions.
The same construction can be applied to other cyclic sectors and seed choices.

The sector labels $(k,q)$ specify the imposed cyclic covariance; additional discrete symmetry may emerge in an individual solution.
For example, the last $(0,7)$ configuration has enhanced $C_{10}$ symmetry, as checked for the full scalar and metric fields.
A rotation by $\pi/5$ leaves the metric invariant and multiplies the scalar by $\ee^{7\pi\ii/5}$.
It therefore also belongs to the $(k,q)=(7,10)$ sector, whose $C_5$ subgroup recovers the imposed $(2,5)$ covariance.
More generally, if $d\ge2$ divides $q$, every solution in sector $(k,q)$ also belongs to sector $(k\bmod d,d)$, since both the metric symmetry and scalar covariance restrict to the subgroup $C_d$.
In particular, the solution space of sector $(k,2k)$ is contained in that of sector $(0,k)$ for $k\ge2$, since the modes $m_s=(2s+1)k$ allowed in $(k,2k)$ form a subset of the modes $m_s=sk$ allowed in $(0,k)$.
The sector labels therefore do not define disjoint families of solutions.

Equal particle number does not imply equal gravitational mass.
Denoting the ADM mass by $M$, we find that the fractional binding $1-M/(\mu N)$ is approximately $(5.69,\,2.80,\,3.85,\,1.59,\,1.57)\times10^{-5}$ for the five bound configurations, ordered from left to right in Fig.~\ref{fig:amplitude}.
We estimate $M$ using the stationary Komar integral, whose asymptotic value equals the ADM mass~\cite{Komar:1958wp,Gourgoulhon:2012ffd}; the global charges are defined in End Matter.

Cyclic symmetry also constrains the ADM angular momentum: $\Rot_q\bm{J}=\bm{J}$ implies $J_x=J_y=0$ for $q\ge2$, so we write $\bm{J}=J\bm{e}_z$.
The five cyclic configurations in Fig.~\ref{fig:amplitude} have noninteger values $J/N\simeq1.601$, $0.2514$, $-1.044$, $3.927$, and $-0.00905$, respectively, whereas conventional axisymmetric stars have integer values.

Time reversal combined with scalar conjugation maps any stationary EKG solution to another at the same positive frequency $\omega$.
For the ansatz~\eqref{eq:stationary}, this transformation acts as $\psi\mapsto\bar\psi$ and $g_{ti}\mapsto-g_{ti}$, leaving $g_{tt}$ and $g_{ij}$ unchanged.
It preserves $M$, $N$, and $|\psi|$, while reversing $J$ and the phase winding around each fixed oriented contour.
Conjugation of the cyclic condition maps $k\mapsto-k\pmod q$, relating the sectors $(k,q)$ and $(q-k,q)$ for $0<k<q$.
Each configuration displayed in the $(2,5)$ sector in Fig.~\ref{fig:amplitude} therefore has a corresponding solution in the $(3,5)$ sector with the same scalar-amplitude distribution and opposite angular momentum.
The sectors with $k=0$ or $q=2k$ are self-conjugate under circulation reversal, but this alone does not require $J=0$.
An individual configuration invariant under circulation reversal, up to a constant scalar phase, must have $J=0$, since reversal changes $J$ to $-J$.
End Matter illustrates this possibility with a cyclic configuration in the $(5,10)$ sector.

\textit{Quantum vortex molecules---}%
The scalar field can be written as $\psi=|\psi|\ee^{\ii S}$ away from its zeros.
Single-valuedness requires an integer winding number
\begin{equation}
    w(C)=\frac{1}{2\pi}\oint_C\dd S\in\mathbb Z,
    \label{eq:winding}%
\end{equation}
for any closed spatial contour $C$ along which $\psi\ne0$.
Nonzero winding around a filament where the scalar field vanishes identifies a quantum vortex~\cite{Fetter:2009zz}.
The cyclic boson stars in Fig.~\ref{fig:amplitude} contain quantized vortex filaments whose relative arrangement remains fixed within a single self-gravitating condensate.
Cyclic symmetry constrains these vortex arrangements.
On a circle centered on the cyclic axis and avoiding scalar zeros, the cyclic condition in Eq.~\eqref{eq:cyclic} fixes the continuous phase change over each angular interval $2\pi/q$ to $2\pi k/q+2\pi s$, with integer $s$.
Continuity keeps $s$ fixed around the circle, so a full turn gives $w=k+sq$.
In a plane perpendicular to the cyclic axis, off-axis vortices form sets of $q$ equally spaced points, with the same winding number within each set~\cite{PhysRevA.79.053820}.
This vortex arrangement motivates their interpretation as quantum vortex molecules.

\begin{figure*}[htbp]
    \centering
    \includegraphics[]{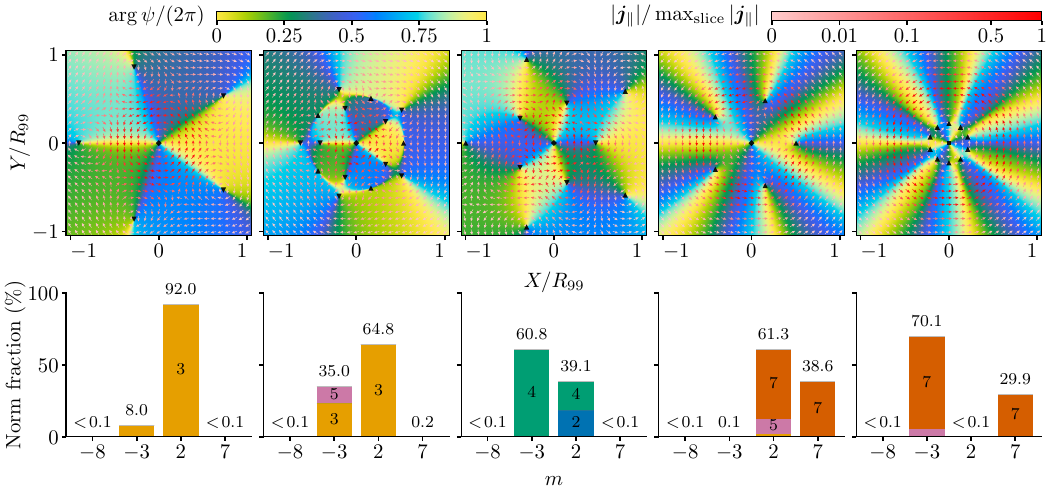}
    \caption{%
    Scalar phase, internal currents, and mode composition of the cyclic boson stars shown in Fig.~\ref{fig:amplitude}, in the same column order.
    Top: scalar phase $\arg\psi/(2\pi)$ at $Z=0.1R_{99}$.
    Arrows indicate the projected current direction; their colors show its magnitude normalized by the maximum magnitude over each displayed slice.
    Markers indicate vortex locations; black upward triangles, downward triangles, diamonds, and squares mark windings $w=+1,-1,+2,-3$, respectively, with counterclockwise winding taken as positive when viewed from $+Z$.
    Bottom: scalar-norm fractions $p_{\ell m}$ stacked at each $m$.
    Colors and labels within the bars identify $\ell$; gray groups individual contributions below $1\%$.
    Numbers above the bars give $p_m=\sum_\ell p_{\ell m}$ in percent.
    }
    \label{fig:phase}
\end{figure*}

The upper panels of Fig.~\ref{fig:phase} show the scalar phase in the slices displayed in Fig.~\ref{fig:amplitude}, with markers indicating the vortex locations.
The first slice contains a central vortex with winding $+2$ and five off-axis vortices with winding $-1$.
The second has a central vortex with winding $+2$, together with ten $-1$ and five $+1$ off-axis vortices.
The third has a central vortex with winding $+2$, together with five $-1$ and five $+1$ off-axis vortices.
The fourth has a central vortex with winding $+2$ surrounded by five $+1$ vortices.
The fifth has a central vortex with winding $-3$ surrounded by ten $+1$ vortices, consistent with its enhanced $C_{10}$ symmetry.
In the displayed slices, the total winding enclosed by zero-free circles centered on the cyclic axis takes the values $2$ or $-3$ for the first three configurations, $2$ or $7$ for the fourth, and $-3$ or $7$ for the fifth, matching their dominant azimuthal mode numbers.

The lower panels of Fig.~\ref{fig:phase} show the azimuthal mode composition underlying these patterns.
Projecting the scalar field onto orthonormal spherical harmonics, we define the coordinate scalar-norm fractions
\begin{equation}
    p_{\ell m}
    =\frac{
        \int r^2\dd r\,
        \left|\int \bar{Y}_{\ell m}(\Omega)\psi(r,\Omega)\dd\Omega\right|^2
    }{
        \int|\psi(\bm x)|^2\dd^3x
    }.
    \label{eq:mode_fraction}%
\end{equation}
These fractions satisfy $\sum_{\ell,m}p_{\ell m}=1$, and summing over $\ell$ gives the azimuthal fraction $p_m=\sum_\ell p_{\ell m}$.
For two azimuthal components $m_1$ and $m_2$, destructive interference produces $|m_2-m_1|$ equally spaced zeros on each circle where their amplitudes match.
The azimuthal separation determines the number of zeros on each equal-amplitude circle, while the spatial profiles determine where these circles occur.
The first three configurations are dominated by the azimuthal components $m=-3$ and $m=2$, whose separation accounts for their fivefold vortex arrangement.
The fourth is dominated by $m=2$ and $m=7$, likewise producing a fivefold arrangement.
The fifth combines $m=-3$ and $m=7$, giving the tenfold arrangement associated with its enhanced $C_{10}$ symmetry.

These configurations support internal scalar currents while their density and vortex geometry remain stationary.
The spatial Noether current measured by observers normal to the time slice is
\begin{equation}
    j^i=\Imo(\bar\psi D^i\psi)=|\psi|^2D^iS,
    \label{eq:internal_current}%
\end{equation}
where $D_i$ is the covariant derivative associated with the spatial metric $\gamma_{ij}$ on a constant-$t$ slice, and $D^i=\gamma^{ij}D_j$.
The phase gradient determines the local flow direction, while the squared scalar amplitude weights the current's strength.
The arrows in the upper panels of Fig.~\ref{fig:phase} show the metric-orthogonal projection of this current onto the displayed slice, as defined in End Matter.

The angular momentum of cyclic boson stars is related to their internal currents and can be understood in terms of their azimuthal mode composition.
In the weak-field approximation, the angular momentum per particle is
\begin{equation}
    \frac{J}{N}
    \simeq\frac{
        \int(\bm x\times\bm j)_z\,\dd^3x
        }{
        \int|\psi|^2\,\dd^3x
        }
    \simeq\sum_m m\,p_m.
    \label{eq:J_N_weak}%
\end{equation}
Here both approximate equalities become exact in the SP limit.
In this limit, $J/N$ is the average of the occupied azimuthal numbers weighted by their norm fractions.
A conventional single-mode axisymmetric star has $J/N=m$, whereas a cyclic star mixes distinct integer $m$ values, whose weighted average is generally noninteger.

The first configuration in Fig.~\ref{fig:phase} has leading norm fractions $92.0\%$ and $8.0\%$ in $m=2$ and $m=-3$, respectively, so the positive angular-momentum contribution dominates, with $J/N\simeq1.601$.
The second has $64.8\%$ and $35.0\%$ in the same modes, producing stronger cancellation, with $J/N\simeq0.2514$.
In the third, the fractions are $60.8\%$ in $m=-3$ and $39.1\%$ in $m=2$, so the negative contribution dominates, with $J/N\simeq-1.044$.
The fourth configuration has leading norm fractions $61.3\%$ and $38.6\%$ in $m=2$ and $m=7$, whose contributions reinforce one another, with $J/N\simeq3.927$.
The fifth has $70.1\%$ and $29.9\%$ in $m=-3$ and $m=7$, producing nearly cancelling contributions, with $J/N\simeq-0.00905$.
These $J/N$ values are obtained from independent ADM evaluations.
Cyclic boson stars thus realize self-gravitating quantum vortex molecules in which quantized vortex winding coexists with generally noninteger angular momentum per particle.

Vortex molecules have been studied in externally confined condensates described by nonrelativistic Gross--Pitaevskii theory~\cite{2008PhRvA..78e3610G}.
In cyclic boson stars, gravity supplies the confinement and supports a stationary cyclic vortex arrangement despite internal circulation.
These EKG solutions provide a relativistic realization of self-gravitating quantum vortex molecules.
The SP limit makes the mechanism supporting these vortex arrangements explicit.
With $U\to0$ at infinity, the Poisson equation in~\eqref{eq:SP} describes an attractive nonlocal interaction with an isotropic $1/r$ kernel.
Interference between azimuthal modes modulates the density, which generates corresponding angular structure in the gravitational potential.
The potential in turn couples the modes, allowing their coherent mixture to satisfy the stationary field equation.
Self-consistent gravity therefore supports stationary discrete rotational order through an isotropic interaction, without external confinement or an imposed angular potential.

The SP equations coincide with the Gross--Pitaevskii--Poisson equations with vanishing contact interaction~\cite{Chavanis:2011zi}.
A laboratory analogue of the nonrelativistic condensate structure of the boson star would require an approximately isotropic $1/r$ attraction over the condensate size, as proposed using off-resonant laser fields~\cite{ODell:1999ewx}.
Coherent superpositions of atomic vortex states have been prepared experimentally~\cite{Andersen:2006evo}.
Preparation could target a coherent mixture of the relevant azimuthal modes with spatial profiles close to those of a stationary cyclic solution, followed by release from the preparation trap while retaining the engineered attraction.
Phase-resolved imaging and angular-mode analysis could then characterize the vortex windings and cyclic symmetry, and test whether the pattern persists without rigid rotation.
Realization requires assessing the achievable interaction range, preparation dynamics, and persistence of the cyclic branches with contact interactions.

\textit{Discussion---}%
We find a new type of stationary boson star with cyclic rotational symmetry in the minimal model of a single massive complex scalar field coupled to gravity.
These solutions exhibit discrete rotational order in self-gravitating condensates and reveal a new way for boson stars to carry angular momentum within this extensively studied model.
Coherent mixtures of azimuthal modes permit generally noninteger angular momentum per particle while preserving quantized vortex winding.
The stationary vortex assemblies provide a relativistic realization of self-gravitating quantum vortex molecules.
Our systematic construction connects weak-field seeds to EKG solutions and provides a framework for exploring these solution branches.
The SP limit shows that an isotropic gravitational interaction can self-consistently support these vortex configurations, motivating possible laboratory analogues in nonrelativistic condensates.

Our focus here is on establishing the existence of cyclic boson stars and identifying their vortex assemblies as self-gravitating quantum vortex molecules.
Other properties, including maximum masses and dynamical stability under perturbations both within and beyond the imposed cyclic sector, remain to be investigated.
The present method constructs EKG solution branches connected to weak-field seeds; more general construction methods and a systematic classification of the solution space remain to be developed.
Extending the construction to include scalar self-interactions or multiple scalar fields would clarify how additional interactions and degrees of freedom affect cyclic boson stars and their vortex configurations.
The formation and subsequent evolution of cyclic boson stars under astrophysical or cosmological conditions remain to be explored.

\textit{Acknowledgments---}%
This work is supported in part by the National Natural Science Foundation of China Grant No.~12588101.

\bibliography{citeLib}

\begin{center}
    \textbf{End Matter}
\end{center}
\appendix

\refstepcounter{section}
\textit{Global charges and projected current---}%
We define the global charges in physical asymptotically Cartesian coordinates $(t,\bm{x})$.
The global $\mathrm{U}(1)$ symmetry gives the conserved Noether charge
\begin{equation}
    N=\int \frac{\ii}{2}\left(\phi\nabla_a\bar{\phi}-\bar{\phi}\nabla_a\phi\right)g^{0a}\sqrt{-g}\,\dd^3x,
    \label{eq:number}%
\end{equation}
interpreted as the particle number, with the integral taken on a constant-$t$ slice.

Write the stationary line element as
\begin{equation}
    \dd s^2=-\alpha^2\dd t^2+\gamma_{ij}(\dd x^i+\beta^i\dd t)(\dd x^j+\beta^j\dd t),
\end{equation}
where $\alpha$, $\beta^i$, and $\gamma_{ij}$ are the lapse, shift, and spatial metric.
With $D_i$ the spatial covariant derivative, our extrinsic-curvature convention is $K_{ij}=(D_i\beta_j+D_j\beta_i)/(2\alpha)$ and $K=\gamma^{ij}K_{ij}$.
On a coordinate sphere $S_r$, let $s^i$ be the outward unit normal within the spatial slice and $\dd A$ its proper area element.
The mass and angular momentum are~\cite{Gourgoulhon:2012ffd}
\begin{align}
    M&=\frac{1}{4\pi}\lim_{r\to\infty}
    \int_{S_r} (D_i\alpha-K_{ij}\beta^j)s^i\,\dd A,
    \label{eq:Komar_mass}%
    \\
    J&=\frac{1}{8\pi}\lim_{r\to\infty}
    \int_{S_r} (K_{ij}-K\gamma_{ij})\varphi^i s^j\,\dd A,
    \label{eq:ADM_spin}%
\end{align}
where $\varphi^i=(\bm e_z\times\bm x)^i$ generates rotations about the cyclic axis at infinity but need not be a Killing field in the interior.
For the stationary, asymptotically flat solutions considered here, with time normalized at infinity, the Komar expression~\eqref{eq:Komar_mass} equals the ADM mass~\cite{Gourgoulhon:2012ffd}.
Equation~\eqref{eq:ADM_spin} defines the total signed ADM angular momentum, including both matter and gravitational contributions.

\begin{figure*}[htbp]
    \centering
    \includegraphics[width=0.329\textwidth]{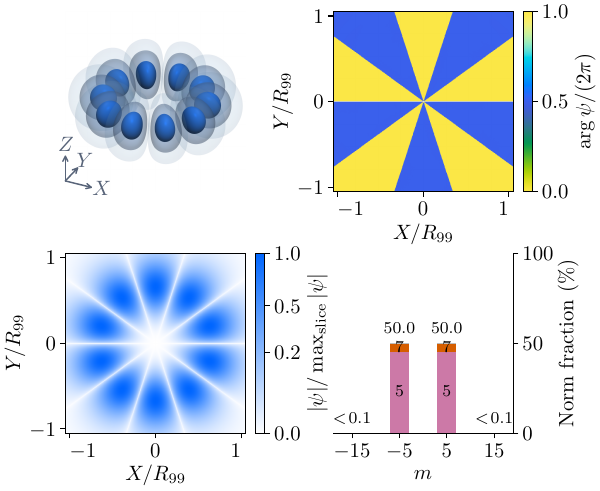}
    \includegraphics[width=0.329\textwidth]{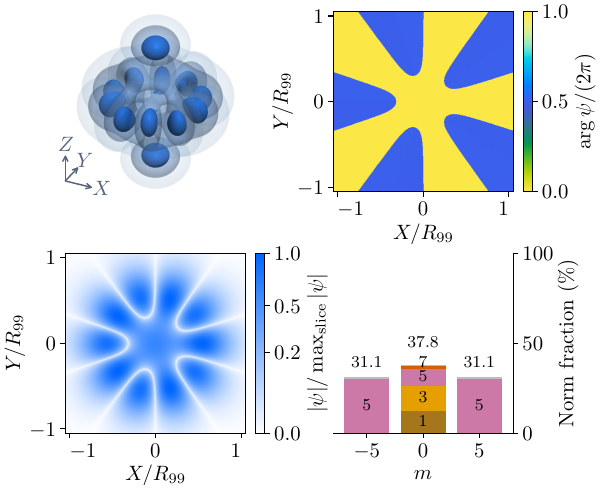}
    \includegraphics[width=0.329\textwidth]{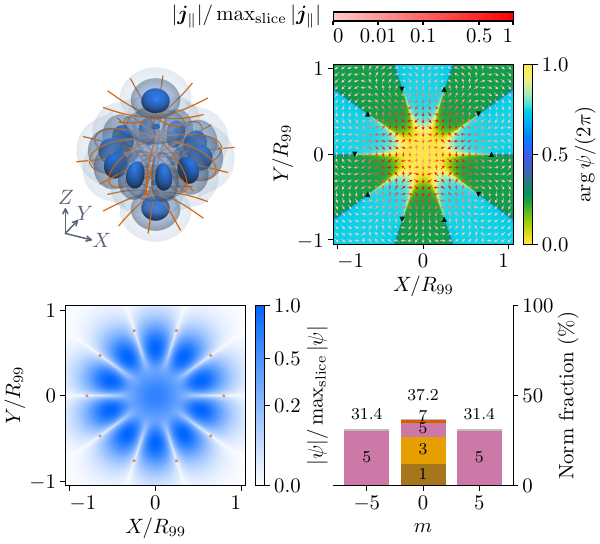}
    \caption{%
    Three cyclic boson stars in sector $(0,5)$ with seed labels $(n,\ell)=(0,5)$ and $\mu^2N=0.1$, all with angular momenta consistent with $J=0$ within numerical accuracy.
    For each configuration, the panels show scalar-amplitude isosurfaces (top left), scalar phase (top right), normalized amplitude (bottom left), and scalar-norm fractions $p_{\ell m}$ (bottom right).
    Plotting conventions follow Figs.~\ref{fig:amplitude} and \ref{fig:phase}.
    The left and middle configurations have negligible internal current; the right supports internal circulation and quantized vortices despite vanishing total angular momentum.
    }
    \label{fig:zero_spin}
\end{figure*}

The current in Eq.~\eqref{eq:internal_current} is measured by observers normal to the time slice.
For Figs.~\ref{fig:phase} and \ref{fig:zero_spin}, it is projected orthogonally onto a plane $z=\mathrm{constant}$ using the spatial metric $\gamma_{ij}$.
Let $\sigma_{AB}=\gamma_{AB}$ be the induced metric on this plane, with $A,B\in\{x,y\}$, and let $\sigma^{AB}$ denote its inverse.
The projected current and its magnitude are
\begin{subequations}
    \begin{align}
        j_\parallel^A&=\sigma^{AB}\Imo(\bar\psi\,\partial_B\psi),
        \\
        |\bm j_\parallel|&=\sqrt{\sigma_{AB}j_\parallel^A j_\parallel^B}.
    \end{align}
\end{subequations}
The constant field and coordinate rescalings used for plotting leave the current directions and normalized magnitudes unchanged.

\refstepcounter{section}
\textit{Cyclic boson stars with $J=0$---}%
Cyclic boson stars can also exist with vanishing total angular momentum.
Figure~\ref{fig:zero_spin} shows three configurations obtained in the imposed sector $(k,q)=(0,5)$ from distinct seeds with the same labels $(n,\ell)=(0,5)$ at $\mu^2N=0.1$.
Their independently evaluated ADM angular momenta are consistent with $J=0$ within numerical accuracy.

The left configuration has enhanced $C_{10}$ symmetry: a rotation by $\pi/5$ leaves the metric invariant and changes the sign of the scalar, so it also belongs to sector $(5,10)$.
This covariance is satisfied by the full scalar and metric fields.
Its spatial scalar profile is real up to a constant phase, and the dominant modes $m=\pm5$ each carry approximately $50.0\%$ of the coordinate scalar norm.
Its ADM angular momentum, $J/N\simeq1.72\times10^{-16}$, is consistent with the symmetry prediction $J=0$ for a configuration invariant under circulation reversal.
It therefore illustrates both the inclusion of sector $(5,10)$ in $(0,5)$ and cyclic structure with vanishing angular momentum in a $(k,2k)$ sector.

The middle configuration also has a real spatial profile up to a constant phase, but contains a substantial $m=0$ component.
The leading norm fractions are $31.1\%$ in each of $m=\pm5$ and $37.8\%$ in $m=0$, with $J/N\simeq-7.43\times10^{-14}$.
The left and middle configurations have negligible shift and spatial current, consistent with static solutions invariant under circulation reversal.
The phase jumps across their nodal sets reflect sign changes of the real scalar profiles and do not represent vortex winding.

Vanishing total angular momentum does not require the spatial current to vanish.
The right configuration has a complex spatial profile and substantial internal current.
Its displayed slice contains five vortices with winding $+1$ and five with winding $-1$, organized in a stationary cyclic arrangement.
The leading norm fractions, $31.4\%$ in each of $m=\pm5$ and $37.2\%$ in $m=0$, are close to those of the middle configuration despite their different local currents.
Its positive and negative angular-momentum contributions cancel, and the independently evaluated ADM value $J/N\simeq3.11\times10^{-9}$ is consistent with zero within numerical accuracy.
This comparison shows that similar azimuthal mode fractions can accompany qualitatively different internal flows, and that stationary quantum vortex configurations can exist with vanishing total angular momentum.

\end{document}